\documentclass[aps,prd,reprint, superscriptaddress, twocolumn]{revtex4-2}

\usepackage{amsmath,amssymb,amsthm,amstext}
\usepackage{natbib}
\usepackage{graphicx}
\usepackage{color}
\usepackage{array, enumerate}
\usepackage{bm}
\usepackage{multirow}
\usepackage[breaklinks,colorlinks,citecolor=blue]{hyperref}
\usepackage{braket}
\usepackage{txfonts}

\def\nn{\nonumber}
\def\be{\begin{equation}}
\def\ee{\end{equation}}
\def\d{{\rm d}}

\begin{document}
\title{Spectral stability of near-extremal spacetimes}
\author{Huan Yang}
\email{hyang@perimeterinstitute.ca}
\affiliation{Perimeter Institute for Theoretical Physics, Ontario, N2L 2Y5, Canada}
\affiliation{University of Guelph, Guelph, Ontario N1G 2W1, Canada}
\author{Jun Zhang}
\email{zhangjun@ucas.ac.cn}
\affiliation{International Centre for Theoretical Physics Asia-Pacific, University of Chinese Academy of Sciences, 100190 Beijing, China}
\affiliation{Taiji Laboratory for Gravitational Wave Universe (Beijing/Hangzhou), University of Chinese Academy of Sciences, 100049 Beijing, China}

\begin{abstract}
It has been suggested that the spectrum of quasinormal modes of rotating black holes is unstable against additional potential terms in the perturbation equation, as the operator associated with the equation is non-self-adjoint. We point out that a bi-linear form has been constructed previously to allow a perturbation analysis of the spectrum, which was applied to study the quasinormal modes of weakly charged Kerr-Newman black holes \cite{Mark:2014aja}. The proposed spectral instability should be restated as instability against potential terms that have infinitesimal ``energy" norm that is specifically defined by the type of inner products introduced by Jaramillo et al. and preserving the physical meaning of energy. We argue that it is necessary to address the stability of all previous mode analysis results to reveal their susceptibility to energetically infinitesimal perturbations. In particular, for near extremal Kerr spacetime, we show that the spectrum of zero-damping modes, which have slow decay rates, is unstable (with order unity fractional change in decay rates) with fine-tuned modification of the potential. The decay rates are, however, always positive with energetically infinitesimal perturbations. If finite potential modifications are allowed near the black hole, it is possible to find superradiantly unstable modes, i.e., a ``black hole bomb" without an explicit outer shell. For the zero-damping modes in near-extremal Reissner-Nordstr\"{o}m-de Sitter black holes, which are relevant for the breakdown of strong cosmic censorship, we find that the corresponding spectrum is stable under energetically infinitesimal perturbations.
\end{abstract}
\maketitle

\section{Introduction}
Modal analysis of black hole spacetimes plays an important role in gravitational-wave astronomy, as the quasinormal mode (QNM) excitation and ringdown contribution is a vital part of generic black hole perturbations, especially for the postmerger black holes. Black hole spectroscopy, i.e., measuring the frequencies and damping rates of QNMs, can be used to infer the black hole mass and spin and to test general relativity. So far, the fundamental mode ($\ell=2,m=2$) has been convincingly detected in some of the LIGO events \cite{LIGOScientific:2018mvr,LIGOScientific:2020ibl,LIGOScientific:2021djp}, including GW150914 \cite{LIGOScientific:2016vlm}. There have been claims of detecting high-overtone modes as well \cite{Isi:2019aib, Capano:2021etf, Capano:2022zqm}, albeit there are concerns from independent analyses \cite{Cotesta:2022pci}. The detection of QNMs with higher $\ell$ generally requires a higher event signal-to-noise ratio, which may benefit from coherently stacking multiple events \cite{Yang:2017zxs}.

It has been recently (re)claimed that distant and/or short-wavelength perturbation of the wave potential of QNMs may significantly change the mode frequencies, despite the perturbation amplitudes being infinitesimal. This observation was initially pointed out in Refs.~\cite{Nollert:1996rf,Nollert:1998ys,Aguirregabiria:1996zy,Vishveshwara:1996jgz} and was recently revisited using a pseudospectrum analysis \cite{Jaramillo:2020tuu,Jaramillo:2021tmt} for the Schwarzschild spacetime. We emphasize that the spectrum of a black hole is stable under infinitesimal perturbations, if the perturbation analysis is performed with the previously constructed bi-linear form in Ref.~\cite{Zimmerman:2014aha}. Instead, the observation in Ref.~\cite{Jaramillo:2020tuu} should be reinterpreted as instability \cite{Gasperin:2021kfv} under energetically infinitesimal perturbations, the norm of which is defined by inner products preserving the physical meaning of energy.

It is crucial to examine the spectral stability under energetically infinitesimal perturbations for various mode analysis results. If the mode frequencies are significantly modified with perturbations in the potential of infinitesimal energy costs, associated claims need to be treated with extra caution due to the susceptibility to external perturbers, or even internal variations due to nonlinearities. A robust claim should have converging measures with respect to small perturbations.
In this work, we focus on the spectral stability of near-extremal black holes, which in general host a class of zero-damping modes (ZDMs) with decay rates approaching zero in the extremal limit \cite{Hod:2008zz,Hod:2008se}. For different kinds of background near-extremal spacetime, these modes have been used to demonstrate parametric nonlinear instability in connection with turbulence \cite{Yang:2014tla}, possible breakdown of strong cosmic censorship (SCC) \cite{Cardoso:2017soq}, and near-horizon critical behavior that leads to the instability of extremal black holes \cite{Gralla:2016sxp,Gralla:2018xzo}, as possibly connected to gravitational critical collapse \cite{Yang:2022okb}. Using the example of near-extremal Kerr, we will discuss whether these modes will be unstable against small perturbations and under what conditions the mode spectrum becomes unstable. We also study the ZDMs for near-extremal Reissner-Nordstr\"{o}m-de Sitter (RNdS) black holes, as an example for nonasymptotically flat spacetimes, and comment on whether the spectral stability affects the divergence on the Cauchy horizon. Throughout the analysis, we adopt the natural unit that $G=c=1$.

\begin{figure}
\includegraphics[height=0.3\textwidth]{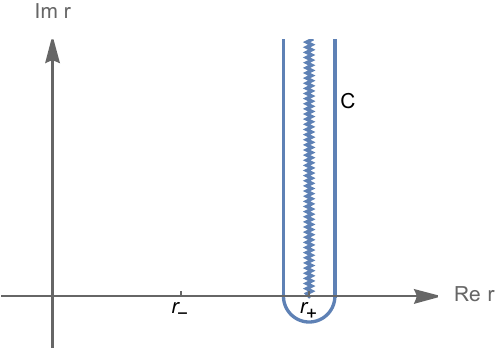}
\caption{\label{fig:cp} An illustration for the contour in the complex $r$-plane that can be used for the perturbation analysis of black hole quasinormal modes. There is a branch cut extending from $r_+$ to infinity. The QNM wavefunctions generally behave as $e^{i\omega r_*}$ and converge to zero at both open ends of ${\cal C}$, making the boundary terms $\langle \chi|H\eta\rangle - \langle H \chi|\eta\rangle$ vanish.} 
\end{figure}

\section{Spectral Stability and Energetic Pseudospectrum}

The eigenvalue spectrum of a non-self-adjoint operator could be unstable under infinitesimal perturbations of the operator, which, in Ref.~\cite{Jaramillo:2020tuu}, is used to explain the significant migration of high-overtone QNM frequencies by small-amplitude yet short-wavelength perturbations of the potential. Nevertheless, perturbative analysis of the spectrum can be made posssible if one replaces the usual inner product with a carefully constructed bi-linear form.
Considering an eigenvalue problem with $H(\omega_0)\, \psi_0=0$ and a perturbation in the potential $H(\omega) \rightarrow H(\omega)+ \epsilon\, \delta V$, we may formally expand the eigenfrequency and eigenfunction as $\omega =\omega_0 +\epsilon\, \omega_1 +\mathcal{O}(\epsilon^2)$ and $\psi =\psi_0+\epsilon\, \psi_1+\mathcal{O}(\epsilon^2)$, respectively. For the unperturbed eigenfunction $\chi$, if one can construct a bi-linear form such that $\langle \chi|H\eta\rangle =\langle H \chi|\eta\rangle$, then we have
\begin{align}
\omega_1 = -\frac{\langle \psi_0 | \delta V |\psi_0 \rangle}{\langle \psi_0 | \partial_\omega H | \psi_0\rangle}\,.
\end{align}
Such bi-linear form is explicitly constructed in Ref.~\cite{Zimmerman:2014aha}. For Kerr black holes it is given by
\begin{align}
\langle \chi | \eta \rangle =\int_{\mathcal{C}} (r-r_+)^s(r-r_-)^s dr \int \sin\theta d \theta \,\chi(r,\theta) \eta(r,\theta)\,,
\end{align}
where $s$ is the spin of the field, $r_\pm$ are the radius of outer and inner horizon, respectively, and $\mathcal{C}$ is the contour as shown in Fig.~\ref{fig:cp}. For example, a Kerr black hole QNM wavefunction can be written as \cite{Leaver:1985ax}
\begin{align}
    R(r) = r^{i \omega r} (r-r_+)^{-s-i\sigma_+}(r-r_-)^{-1-s+i \omega +i \sigma_+} \sum^\infty_{n=1} d_n \left( \frac{r-r_+}{r-r_-}\right)^n
\end{align}
with $\sigma_+ =(\omega r_+-a m)/(r_+-r_-)$. The $(r-r_+)^{-s-i\sigma_+}$ term, which is associated with the horizon boundary condition, leads to a branch cut starting at $r=r_+$ as we analytically continue the wavefunction to the complex r plane. The other branch cut starting at $r=r_-$ is not relevant for the discussion here.
Having in mind $H(\omega)$ taking the form of
\begin{align}
H(\omega)=\Delta^{-s}\frac{d}{dr} \left (\Delta^{s+1} \frac{d}{dr} \right ) - V(\omega,r)
\end{align}
with ($\Delta=(r-r_+)(r-r_-)$), the boundary terms are $\Delta^{s+1}\left( \chi \frac{\d \eta}{\d r} -  \frac{\d \chi}{\d r} \eta \right )_{\bar{\mathcal C}}$. Here $\bar{\mathcal C}$ denotes the open ends of the contour. The QNM wave functions generally have $e^{i \omega r}$ behavior for large $r$~\cite{Leaver:1985ax}, where the mode frequency $\omega$ has a positive real part. It is straightforward to see that the mode wave function quickly converges to zero at  both open ends of $\mathcal{C}$, and the boundary terms vanish identically.  
The corresponding eigenvalue analysis is applied for computing the QNM frequency of weakly charged Kerr-Newman black holes \cite{Mark:2014aja}, showing no instability of spectrum. The results obtained by the perturbative analysis in Ref.~\cite{Mark:2014aja} were later shown to be fully consistent with the numerical quasinormal mode frequencies in Ref.~\cite{Dias:2015wqa}. Also see Ref.~\cite{Green:2022htq} for a construction of bi-linear form with the focus on the orthogonality of QNMs.

The high-frequency and the distant Gaussian-bump perturbations considered in Refs.~\cite{Jaramillo:2020tuu,Cheung:2021bol} can all be consistently treated within the above formalism, although the normalizations of these operators are dramatically different. In fact, similar perturbation analysis has been performed in Ref.~\cite{Leung:1997was} for a scalar field in Schwarzschild spacetime, in which case the mode frequency perturbation is shown to scale as $e^{2 i \omega_0 x_V}/x^2_V$ for a bump centered around $x_V$.  Since ${\rm Im}(\omega_0)<0$, the importance of the potential perturbation is exponentially amplified by its distance $x_V$. Moreover, this formalism produces useful insights for results from pseudospectrum analysis. Assuming the QNM frequency shift is $\omega_1(r_V)$ for a potential perturbation $\epsilon\delta(r-r_V)$, the frequency shift would be $\epsilon\int dr_V \omega_1(r_V) \delta V(r_V)$ for a general perturbation $\epsilon \delta V(r)$, as long as $\epsilon$ is controlled to ensure a small shift. For $\delta V \sim e^{i k r}$ as discussed in Ref.~\cite{Jaramillo:2020tuu}, the integral contributed by potential perturbations near the horizon scales as $k^{-2 {\rm Im}(\omega)-1}$ as $\omega_1(r) \propto \psi^2_0 \propto (r-r_+)^{-2i \omega}$ near the horizon at $r_+$. So high-overtone modes are more susceptible to high-$k$ perturbations.

Despite not making the operator self-adjoint, the inner product defined in Ref.~\cite{Jaramillo:2020tuu} closely fits the intuitive expectation of the magnitude (or normalization) of a potential. The phenomena observed are physically relevant. To reconcile with the above discussion and to be precise with the terminology, we shall follow the convention to refer the analysis in Ref.~\cite{Jaramillo:2020tuu}  as ``the energetic pseudospectrum" to emphasize the special choice of inner product \cite{Gasperin:2021kfv}. For a generic mode analysis, it is necessary to address the spectral stability under energetically small perturbations, as it characterizes the robustness of the mode spectrum and the associated implications, such as mode stability and SCC. In the following, we shall consider distant perturbations as examples of energetically small perturbations, and investigate the migration of the ZDMs of near-extremal Kerr and RNdS black holes.

\section{Spectral stability of near-extremal Kerr}

For Kerr black holes, ZDMs start to emerge when the dimensionless spin of black hole $a$ becomes greater than a certain critical value \cite{Yang:2012pj}, and in the following we shall focus on the near-extremal case such that $\epsilon_a\equiv1-a \ll1$. In order to address the spectral stability of the ZDMs, we shall consider a bosonic field in the Kerr spacetime, and investigate the frequency migration using a matched expansion analysis.  It consists of finding
an exact solution in a region near spatial infinity, near the horizon and matching the expressions in between.

It is known that perturbations of the bosonic field can be described by a master variable ${}_s \psi = R(r)S(\theta)e^{i m \phi-i \omega t}$ which satisfies the separable Teukolsky equations. Here $s$ is the spin weight of the bosonic field. The ZDM frequencies are the eigenfrequencies of the radial Teukolsky equation 
\be\label{eqLR}
\mathcal{L}\left(s,\,\lambda,\,\omega\right) R=0\, ,
\ee 
where $\lambda ={}_s A_{\ell m \omega}+\omega^2-2 m \omega$ with ${}_s A_{\ell m \omega}$ being the eigenvalue of the angular Teukolsky equation. 

It is convenient to define a dimensionless radial coordinate $x \equiv r/M-1$ with $M$ being the black hole mass. To perform the matched expansion analysis, we first consider the exterior regime ($x \gg \sqrt{\epsilon_a}$), where Eq.~\eqref{eqLR} reduces to
\be\label{eqTeuR}
R''(x)+\frac{2(s+1)}{x}R'(x)+\frac{\omega^2(x+2)^2+2i\omega s x -\lambda}{x^2}R(x) =0,
\ee
for $\epsilon_a \ll 1$. The homogeneous solutions of Eq.~\eqref{eqTeuR} can be expressed as
\be\label{eq:far}
R(x \gg \sqrt{\epsilon_a}) = A F_+(x)+B F_-(x)
\ee
with
\be
F_\pm(x)= e^{-i \omega x} x^{-\frac12-s\pm i\delta}\, {}_{1}F_1\left(\frac12+s\pm i \delta+2i \omega, 1\pm2 i \delta, 2 i \omega x\right)\,,\nn
\ee
where ${}_{1}F_1(z)$ is the confluent hypergeometric function and $\delta^2 \equiv7m^2/4-(s+1/2)^2-{}_s A_{\ell m \omega}$. Imposing the outgoing boundary condition at infinity leads to the physical solution
\be
R_{\rm o} \left(x \gg \sqrt{\epsilon_a}\right) = A_{\rm o} F_+(x) + B_{\rm o} F_-(x)
\ee
with
\begin{align}\label{eq:farratio}
\frac{A_{\rm o}}{B_{\rm o}} = e^{\pi \delta + 2 i \delta \log 2\omega} \frac{\Gamma(-2i\delta) \Gamma(1/2-s+i \delta -2 i \omega)}{\Gamma(2 i \delta) \Gamma(1/2-s-i \delta -2 i \omega)}\,.
\end{align}

Now we introduce a distant perturbation in the exterior regime, and the perturbed radial equation can be written as
\begin{align}
\mathcal{L}\, R+\epsilon\, \delta V(x)\, R=0\, ,
\end{align}
where the support of $\delta V(x)$ is compact in the exterior regime. The perturbation modifies the radial function: To the first order in $\epsilon$, we have $R = \tilde{R} + {\cal O}(\epsilon^2)$ with 
\begin{align}
 \mathcal{L}\, \tilde{R}=-\epsilon \delta V(x) R_{\rm o}\,.
\end{align}
To evaluate the modified radial function $\tilde{R}$, let us consider a Green function $g(x,x')$ which satisfies
\begin{align}\label{eq:green}
\mathcal{L}\, g(x,x')=\delta(x-x')\, ,
\end{align}
and the outgoing boundary condition at infinity. 
For $x \neq x'$, the Green function $g(x, x')$ in the exterior regime can be described by the homogenous solution~\eqref{eq:far}. For $x>x'$ the coefficients $A$ and $B$ are $A_{\rm o}$ and $B_{\rm o}$ satisfying Eq.~\eqref{eq:farratio} given the outgoing boundary condition, while for $x<x'$ the coefficients, say, $A_{\rm in}$ and $B_{\rm in}$,  can be obtained by matching $g(x,x')$ at $x=x'$.
In particular, according to Eq.~\eqref{eq:green}, the value of $g$ is continuous at $x=x'$, leading to 
\begin{align}
 A_{\rm in}F_+(x') + B_{\rm in}F_-(x') =  A_{\rm o}F_+(x') + B_{\rm o}F_-(x')\, ,
\end{align}
and the derivative of $g'$ satisfies
\begin{align}
 A_{\rm in}F'_+(x') + B_{\rm in}F'_-(x') +1=  A_{\rm o}F'_+(x') + B_{\rm o}F'_-(x')\,.
\end{align}
Hence, we find
\begin{align}
A_{\rm in} =A_{\rm o}-1/W(x') \quad  B_{\rm in}=B_{\rm o}+1/W(x')\, ,
\end{align}
where $W(x) \equiv F_+(x)F'_-(x)-F_-(x)F'_+(x) $ is the Wronskian of the two homogeneous solutions. 
The modified radial function, using the Green's function, is
\begin{align}
\tilde{R} =R_{\rm o}(x)-\epsilon \int d x' g(x,x') \delta V(x') R_{\rm o}(x')\,.
\end{align}
For $x$ on the right side of the support of $\delta V$, $\tilde{R}$ is given by
\begin{align}
\tilde{R}_> 
\equiv \left [1-\epsilon \alpha\right ] \times R_{\rm o}(x)\,,
\end{align}
and for $x$ on the left side of the support of $\delta V$, $\tilde{R}$ is given by
\begin{align}
\tilde{R}_< 
\equiv  R_>+\epsilon \beta  [F_+(x)-F_-(x)]\,
\end{align}
where
\begin{align}
\alpha  \equiv \int dx' \delta V(x') R_{\rm o}(x') \quad \beta  \equiv \int dx' \delta V(x') \frac{R_{\rm o}(x')}{W(x')}\,.
\end{align}

In order to obtain the eigenfrequencies, we need to extend the exterior solution to the black hole horizon. This can be done by matching the exterior solution to the interior solution in the intermediate regime. Specifically, the radial equation in the interior regime ($x\ll1$) can also be simplified, and the solutions can be expressed by hypergeometric functions~\cite{Yang:2013uba}. In the intermediate regime where $x \ll 1$ but $x\gg \sqrt{\epsilon_a}$, both the interior and exterior solutions can be expressed as linear combinations of $x^{-1/2-s+i \delta}$ and $x^{-1/2-s-i \delta}$. 
In particular, in the intermediate regime, the modified exterior radial function behaves like
\begin{align}
&\tilde{R}_< \, \left(\sqrt{\epsilon_a}\ll x \ll 1\right)  \nn \\
&\rightarrow \left(A_{\rm o} -\epsilon \alpha A_{\rm o} +\epsilon \beta\right)x^{-\frac12-s+i \delta}+\left(1 -\epsilon \alpha  -\epsilon \beta\right) x^{-\frac12-s-i \delta},
\end{align}
where we have set $B_{\rm o}=1$ without loss of generality. The interior and exterior solutions can be matched by identifying the ratio of the coefficients of of $x^{-1/2-s+i \delta}$ and $x^{-1/2-s-i \delta}$, leading to the equation for the eigenfrequency,
\begin{widetext}
\begin{align}
e^{-\pi \delta-2 i \delta \ln(m)-i\delta \ln(8\epsilon)}\frac{\Gamma^2(2 i \delta)\Gamma(1/2+s-i m -i \delta)\Gamma(1/2-s-i m -i \delta)\Gamma[1/2+i(m-\delta-\sqrt{2}\bar{\omega})]}{\Gamma^2(-2 i \delta)\Gamma(1/2+s -i m +i \delta)\Gamma(1/2-s-i m +i \delta)\Gamma[1/2+i(m+\delta-\sqrt{2}\bar{\omega})]} =\Pi\,.
\end{align}
\end{widetext}
Here, $\Pi$ is defined by the ratio from the exterior solution
\begin{align}\label{eqAPi}
A_{\rm o} \Pi \equiv A_{\rm o} \frac{1-\epsilon \alpha +\epsilon \beta/A_{\rm o}}{1 -\epsilon \alpha  -\epsilon \beta}\, ,
\end{align}
and $\bar{\omega} \equiv (\omega-m \Omega_{\rm H})/\sqrt{\epsilon_a}$ with $\Omega_{\rm H}$ being the horizon frequency.

In the absence of perturbation $\delta V$, we have $\Pi=1$, and the ZDM frequency is obtained by noticing that $\Gamma[1/2+i(m-\delta-\sqrt{2}\bar{\omega})]$ is near its pole \cite{Yang:2013uba}, as the rest of the terms multiplied together is rather small. To understand the frequency migration caused by $\delta V$, it is therefore instructive to look at the magnitude of $\Pi$ in more detail. We shall, for example, take a narrow Gaussian $\delta V$ centered around $x_V$ with $\int dx \,\delta V=1$, which is capable of significantly shifting the fundamental mode of the near-extremal Kerr. In this case, we have
\begin{align}
\alpha & \approx R_{\rm o}(x_V),\quad  \beta   \approx R_{\rm o}(x_V)/W(x_V)\,.
\end{align}
The dominant amplitude growth in $R_{\rm o}$ comes from the $e^{i \omega x}$ factor. In the case that ${\rm Re}(i \omega x) \ll \mathcal{O}(\log 1/\epsilon)$, or $x\ll \mathcal{O}(1/\sqrt{\epsilon_a}\log 1/\epsilon)$, the $\epsilon R_{\rm o}(x)$ term is small, so that $\Pi$ is close to one.   In order to have significant impact on $\omega$, we need $\Pi \approx 0$. This can be achieved only by rather fine-tuned $x_V$ and $\epsilon$. We will now focus on this case and investigate whether it possibly allows unstable modes.
The requirement of $\Pi \approx 0$ becomes 
\begin{align}\label{eq:1oep}
R_{\rm o}(x_V)\left [ 1-\frac{1}{W(x_V)A_{\rm o}}\right ] \approx \frac{1}{\epsilon}\,.
\end{align}

\begin{figure}
\includegraphics[height=0.4\textwidth]{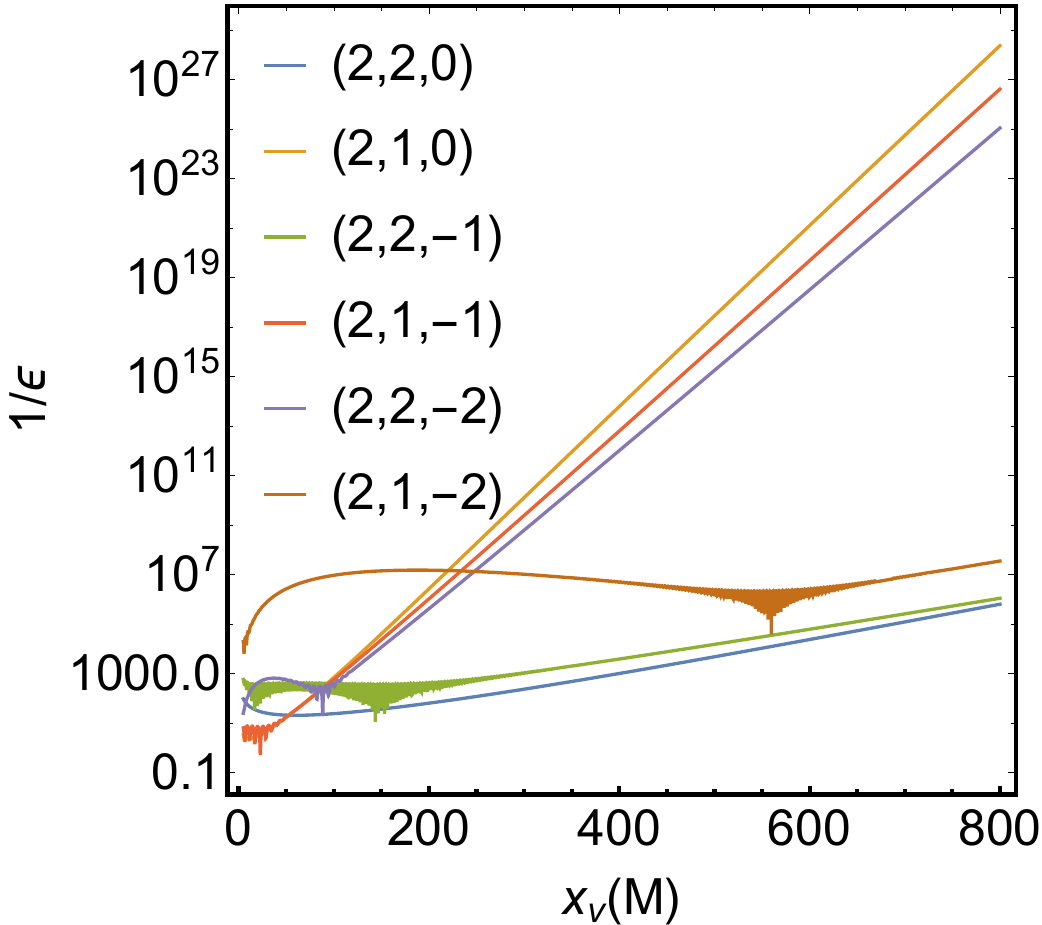}
\caption{\label{fig:ep}  The magnitude of the lhs of Eq.~\eqref{eq:1oep} as a function of the radial position $x_V$, for modes with different $(\ell,m,s)$. The exponential divergence of the wave function at spatial infinity amplifies the effect of a distant, small potential perturbation. } 
\end{figure}

If the decay rate is positive, for any infinitesimal $\epsilon$, we can also find $x_V$ such that the solution exists (as illustrated in Fig.~\ref{fig:ep}), although the values are rather fine tuned, as $R_{\rm o}$ is oscillatory. If the decay rate is negative or zero, $R_{\rm o}(x_V)$ (notice that $R_{\rm o} \sim e^{i \omega x} /x^{1/2+i \delta-s-2 i \omega}$ for $x \gg 1$) and the lhs of Eq.~\eqref{eq:1oep} are bounded in magnitude for all $x_V$, so unstable modes are not allowed for infinitesimal $\epsilon$. In fact, for more general $\delta V$, if the mode decay rate is positive, the corresponding $\alpha$ and $\beta$ are all bounded as $R_{\rm o}$ is bounded, so that $\Pi \approx 0$ cannot be realized by infinitesimal $\epsilon$, indicating that ZDMs are always stable by energetically infinitesimal perturbations.

\begin{figure}
\includegraphics[height=0.4\textwidth]{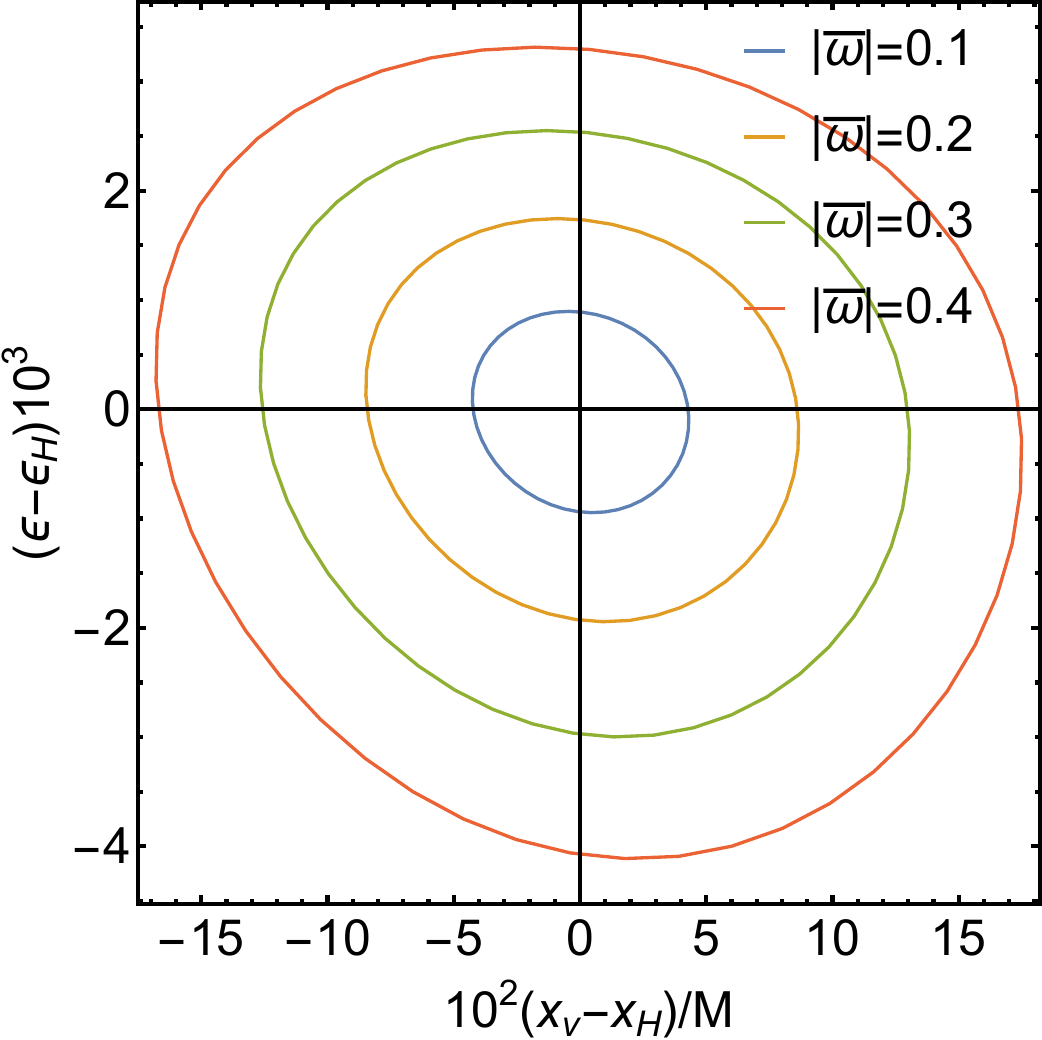}
\caption{\label{fig:con}  The amplitude $\epsilon$ and position $x$ of a $\delta$-type potential perturbation placed near the black hole, in order to match a ZDM with rescaled frequency $\bar{\omega}$. As each contour labeled with fixed $|\bar{\omega}|$ maps to a circle in the complex $\bar{\omega}$ plane with the same fixed $|\bar{\omega}|$, part of the contour corresponds to unstable modes.} 
\end{figure}

However, for finite (and large enough) $\epsilon$, it seems we still can find $x_V$ to have Eq.~\eqref{eq:1oep} satisfied, so that an unstable mode is allowed. For example, for scalar modes with $(\ell,m)=(2,2)$ and a $\delta$-type potential $\epsilon \delta(x-x_V)$, we find that $(x_{\rm H}/M,\epsilon_{\rm H}) \approx (7.26,-0.0178)$ gives rise to $\omega= m \Omega_{\rm H}$. For modes with modified frequencies and decay rates, the corresponding fine-tuned position and amplitude are shown in Fig.~\ref{fig:con}. Part of the modes there are unstable. This kind of realization is different from putting a reflective boundary on a finite radius that has been proposed, to make a black hole bomb \cite{Press:1972zz,Cardoso:2004nk}.
It will be interesting to investigate whether the unstable mode similar to the one shown in Fig.~\ref{fig:con} is still allowed with the black hole spin away from unity, for small but not infinitesimal perturbations.


\section{Spectral Stability of near-extremal RNdS}
A RNdS black hole possesses three horizons: the Cauchy horizon, event horizon, and cosmological horizon, from inside to outside, located at $r_-$, $r_+$, and $r_c$, respectively. Considering a minimally coupled scalar field on a RNdS background, Ref.~\cite{Cardoso:2017soq} shows that the mode spectrum of the scalar field allows universal ``fast enough" decay rates, so that the initial data are sufficiently regular on the Cauchy horizon to violate the SCC. Note that the nonasymptotic flatness is crucial for the argument here, as the spacetime is free of the power-law tail that introduces the mass-inflation effect \cite{Poisson:1990eh}. In the near-extremal case [$\tau \equiv (r_+ - r_-)/r_+ \rightarrow 0$], the mode with slowest decay rate, namely, the dominating mode, belongs to a set of purely decaying modes. The frequencies of such modes behave as
\be\label{wNE}
\omega_{\rm NE} \rightarrow - i (l+n+1)\kappa_+ \, ,
\ee
as $\tau \rightarrow 0$ independent of $r_c$ \cite{Cardoso:2017soq}. Here $\kappa_+$ is the surface gravity at the event horizon. As the magnitude of $\omega_{\rm NE}$ decides the regularity of the Cauchy horizon, it is important to show whether the values are susceptible to small perturbations of the RNdS spacetime.

It turns out that the matched expansion analysis performed in the Kerr case is difficult to implement in the RNdS spacetime as the perturbation equation does not simplify to yield analytical solutions in all relevant domains. Nevertheless, similar to the Kerr analysis, it is instructive to first check the effect of potential perturbation on the mixing of homogeneous solutions in the region away from event horizon, and then address how the mode frequency migrates under this mixing. As in the Kerr case, QNM wave functions should be bounded away from horizons, so that  potential perturbations that possibly affect the spectral stability can appear only near the cosmological horizon, which we focus on below.

In the region near the cosmological horizon, the radial function of the dominating mode can be described by 
\be\label{eqRxc}
R(x\rightarrow 1/c) =A_{\rm o} F_+(x) + B_{\rm o} F_-(x)
\ee
with $F_\pm (x) = \Gamma(1\mp 2iw) {\rm J}_{\mp 2 i w}\left(2j \sqrt{1- cx}\right)$. Here, ${\rm J}_n(z)$ is the Bessel function of the first kind, and we define $x\equiv r/r_+$, $c \equiv r_+/(r_c-r_+)$,
\be
j^2 \equiv \frac{(1+4 c+6 c^2)}{2+4 c}\ell(\ell+1)\,\quad {\rm and} \quad w \equiv \frac{\omega}{2 \kappa_c}
\ee
with $\kappa_c$ being the surface gravity at $r_c$. The outgoing boundary condition indicates $A_{\rm o}/B_{\rm o}=0$ in the absence of perturbations.

In the presence of a potential perturbation $\delta V = \epsilon \delta (x-x_V)$, the corrections on $A_{\rm o}/B_{\rm o}$ can be obtained in a similar way as in the Kerr case [cf. Eq.~\eqref{eqAPi}]:
\be\label{ABP}
\frac{\tilde{A}_{\rm o}}{\tilde{B}_{\rm o}} = \frac{(1-\epsilon \alpha) A_{\rm o} - \epsilon \beta}{(1-\epsilon \alpha) B_{\rm o} + \epsilon \beta} \approx \frac{A_{\rm o}}{B_{\rm o}} - \epsilon \frac{A_{\rm o}+B_{\rm o}}{B_{\rm o}^2}\frac{R_{\rm o}(x_V)}{W(x_V)}\, .
\ee
In particular, we find that the correction on $A_{\rm o}/B_{\rm o}$ is of the order of $\epsilon (1 - c x_V)^{1+i \omega/2\kappa_c}$ (see App.~\ref{app:RNdS} for more details). Considering $\omega \sim \tau$, we conclude that the correction on $A_{\rm o}/B_{\rm o}$ due to an infinitesimal perturbation in the potential is bounded.

\begin{figure}[tp]
\centering
\includegraphics[height=0.21\textwidth]{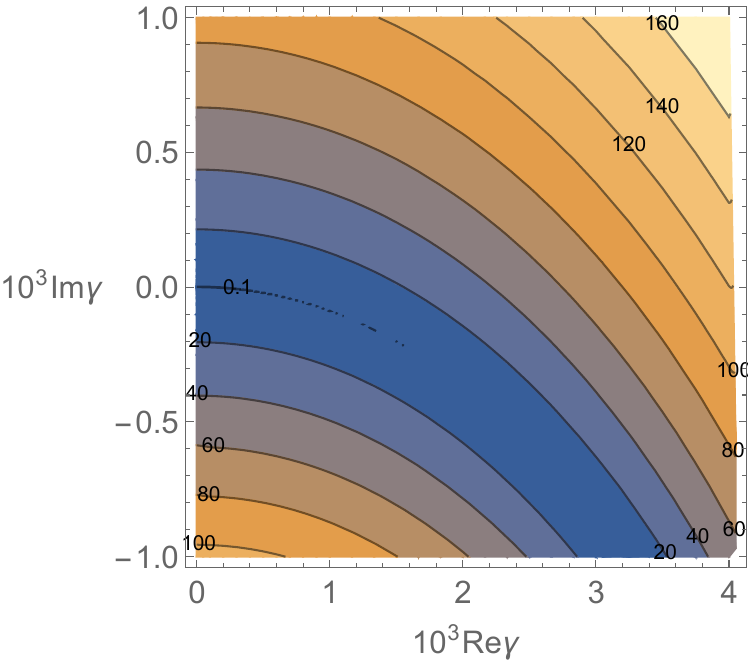}
\includegraphics[height=0.21\textwidth]{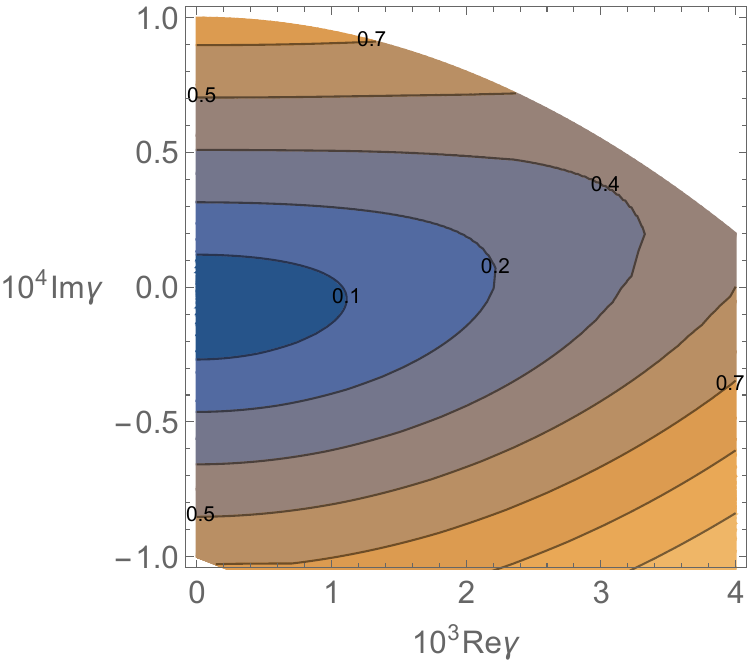}
\caption{\label{fig:ratio}The ratio between coefficients of the two independent solutions $F_\pm$ to the equation of the radial function. The ratio shows how the two solutions $F_\pm$ get mixed due to the presence of distant potential perturbations. The plots show the cases with $\Lambda M^2 = 0.06$ (left)  and  $\Lambda M^2 = 0.2$ (right), and $Q=0.998 Q_{\rm Max}$ in both cases. $\gamma = (\omega-\omega_{\rm NE})/|\omega_{\rm NE}|$ is the relative deviation of $\omega_{\rm NE}$, the numbers on the contours show $10^5 \times |\tilde{A}_{\rm o}/\tilde{B}_{\rm o}|$, and $F_{\pm}$ is chosen so that the ratio in the absence of perturbations $\tilde{A}_{\rm o}/\tilde{B}_{\rm o}=0$.} 
\end{figure}
\begin{figure}[tp]
\centering
\includegraphics[scale=0.6]{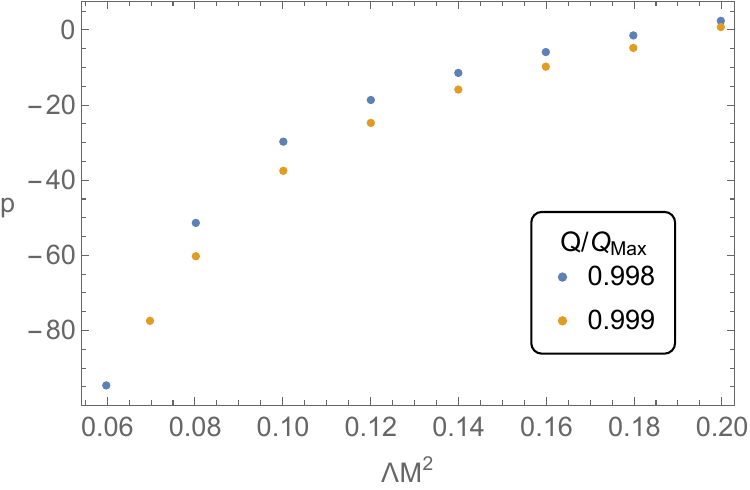}
\caption{$p$ as a function of $\Lambda M^2$. Here, $p$ is the parameter in ${\rm Im}\gamma = p\, \left({\rm Re}\gamma\right)^2 + q$ which described the blue ``valley" in Fig.~\ref{fig:ratio}. Note that $\Lambda M^2$ is bounded by $\Lambda_{\rm Max}M^2 \sim 0.2$ for the $Q$ considered here, beyond which the cosmological horizon would be smaller than the event horizon \cite{Cardoso:2017soq}. \label{fig:p}} 
\end{figure}

To further understand the effects of perturbations on the QNM frequencies, we need to address how the mode frequencies change with $\tilde{A}_{\rm o}/\tilde{B}_{\rm o}$. In practice, we numerically solve the radial wave equation of different $\omega$ with an ingoing boundary condition at the event horizon, and then extract $\tilde{A}_{\rm o}/\tilde{B}_{\rm o}$ by matching the numerical solution with Eq.~\eqref{eqRxc} near the cosmological horizon. Sample numerical results are shown in Fig.~\ref{fig:ratio}, where we introduce $\gamma = (\omega-\omega_{\rm NE})/|\omega_{\rm NE}|$ to describe the relative deviation from $\omega_{\rm NE}$, and the contours show the absolute value of $\tilde{A}_{\rm o}/\tilde{B}_{\rm o}$. We expect $\tilde{A}_{\rm o}/\tilde{B}_{\rm o}=0$ in the absence of perturbation, which is indeed what we find at $\gamma = 0$ up to numerical errors. As shown in Fig.~\ref{fig:ratio}, the ZDM frequency is stable against infinitesimal perturbations. This is especially the case for relatively larger $\Lambda$. For small $\Lambda$, it is possible to fine-tune the perturbation so that it can modify the frequency at a relatively small cost. Namely, by fine-tuning the perturbations, the frequency could drift along the blue ``valley" in the left plot in Fig.~\ref{fig:ratio}. The cost becomes smaller as $\Lambda$ decreases, and eventually we expect the frequency to become unstable in the limit of $\Lambda \rightarrow 0$. This pattern is generic. Actually, the position of the valley can be numerically fitted by ${\rm Im}\gamma = p \left({\rm Re}\gamma\right)^2 + q$, where $q$ is tiny and affected only by numerical errors. As shown in Fig.~\ref{fig:p}, $p$ increases as $\Lambda$ decreases, and we expect it to approach $-\infty$ as $\Lambda$ goes to zero.

In fact, the QNM frequencies in the limit of $\Lambda \rightarrow 0$, i.e., the near-extremal RN black hole, have been studied in Ref.~\cite{Hod:2017gvn}. Following a similar matched expansion analysis as in the Kerr case and in Ref.~\cite{Hod:2017gvn}, we find that the correction on $A_{\rm o}/B_{\rm o}$ is proportional to $x_V^{1 - 2 i \omega r_+} e^{i \omega r_+ x}$, which is unbound as $x_V$ goes to infinity. Therefore, even with an infinitesimal perturbation in potential (maybe fine-tuned), the frequencies of the ZDMs can change significantly.

Regarding the SCC, our results also show that infinitesimal perturbations only infinitesimally modify the frequencies. We expect the decay rate of the scalar perturbations on the extremal RNdS spacetime is still sufficiently fast for the violation of the SCC~\cite{Cardoso:2017soq} even in the presence of perturbations. Figure~\ref{fig:p} actually indicates that the decay rates tend to be larger with potential perturbations. 
\\

\section{Discussion}
The analysis for perturbations in near-extremal Kerr and RNdS spacetimes indicates that the spectrum is stable for near-extremal spacetimes with nonzero cosmological constants, and unstable with respect to fine-tuned small perturbations if the cosmological constant is zero.  It is not explicitly shown but nevertheless reasonable to expect that this applies for general Kerr-Newman-type black holes. On the other hand, it seems the infinitesimal perturbations on the potential do not change the stability of modes, nor do they influence the robustness of the claims regarding the SCC. Of course these claims ought to be explicitly checked in the relevant spacetimes.

It will also be interesting to investigate the shift of time domain signals in response to a small modification in the potential. The ZDMs of extremal spacetimes are known to excite collectively by external sources \cite{Yang:2013uba}, giving rise to (transient) power-law signals. In the presence of potential perturbations, the time domain signal may exhibit other intriguing behavior  beyond the understanding for individual modes. \\

{\it Acknowledgements---} 
We thank Kyriakos Destounis for help with computing modes for RNdS  black holes. We also thank the anonymous referee for the useful comments on the manuscript. J. Z. is supported by the scientific research starting grants from University of Chinese Academy of Sciences (Grant No.~118900M061), the Fundamental Research Funds for the Central Universities (Grant No.~E2EG6602X2 and Grant No.~E2ET0209X2), and the National Natural Science Foundation of China (NSFC) under Grant No. 12147103. H. Y. is supported by the National Science and Engineering Research Council through a Discovery grant. This research was supported in part by Perimeter Institute for Theoretical Physics. Research at Perimeter Institute is supported by the Government of Canada through the Department of Innovation, Science and Economic Development Canada and by the Province of Ontario through the Ministry of Research, Innovation and Science.

\bibliography{ms}

\appendix

\section{QNMs of a massless scalar field in the near-extremal RN(dS) spacetime}\label{app:RNdS}

In this appendix, we discuss the QNMs of a massless scalar field in the near-extremal RN(dS) spacetime. The metric of the RNdS spacetime can be written as
\be\label{metricRNdS}
\d s^2 = - f(r) \d t^2 + \frac{1}{f(r)} \d r^2 + r^2 \left( \d \theta^2 + \sin^2\theta \d \phi^2 \right) \, 
\ee
where
\be
f(r) = 1 - \frac{2 M}{r} + \frac{Q^2}{r^2} - \frac{\Lambda r^2}{3} 
\ee
with $M$, $Q$, and $\Lambda$ being the BH mass, BH charge, and the cosmological constant, while the metric RN spacetime is given by Eq.~\eqref{metricRNdS} with $\Lambda=0$. A RNdS black hole possesses three horizons. From inside to outside, they are the Cauchy horizon, event horizon, and cosmological horizon, and are located at $r_-$, $r_+$, and $r_c$, respectively. We introduce $\kappa_{\pm, c} = \left| f'(r_{\pm, c})/2\right|$ to denote the surface gravity of each horizon. 

The equation of the massless scalar field is
\be\label{KGeq}
\Box\, \Phi = 0
\ee
where $\Box$ is the d'Alembert operator on the RNdS spacetime. Substituting
\be
\Phi = \int \d \omega\, \sum_{\ell m}\, e^{-i \omega t}\,   R_{\omega \ell m}(r)\, Y_{\ell m}(\theta)\, e^{i m \phi} 
\ee
into Eq.~\eqref{KGeq} leads to the equation of radial function
\be\label{eqR}
f(r) \frac{\d}{\d r} \left[ r^2 f(r) \frac{\d R}{\d r} \right]+ U(r)\, R = 0\, ,
\ee
where
\be
U(r) = \omega^2 r^2 - \ell(\ell+1) f(r)\, .
\ee

\subsection{RN spacetime}
We shall start with the near-extremal RN spacetime by closely following and summarizing some results of Ref.~\cite{Hod:2017gvn}, where the QNMs of a scalar field on the near-extremal RN spacetime have been studied. Introducing the dimensionless parameters
\be
x \equiv \frac{r-r_+}{r}, \quad \tau \equiv \frac{r_+-r_-}{r_+}, \quad k \equiv 2\omega r_+, \quad \varpi \equiv \frac{k}{\tau},
\ee
and
\be
\beta^2 = (\ell+1/2)^2-k^2,
\ee
Eq.~\eqref{eqR} can be written as
\be\label{eqRx}
x(x+\tau) \frac{\d^2 R}{\d x^2} + (2 x + \tau) \frac{\d R}{\d x} + U(x)\, R = 0
\ee
where
\be
U(x) = \frac{\left(\omega r_+ x^2 + k x + \varpi \tau/2\right)^2}{x(x+\tau)}- \ell(\ell+1)\,.
\ee
We assume $0 < \beta \in \mathbb{R}$.

In the near region, i.e., $x\ll1$, we have $U \rightarrow  \left(k x + \varpi \tau/2\right)^2/\left[x(x+\tau)\right]- \ell(\ell+1)$, and the physical solution, i.e. the one satisfies the ingoing boundary condition, is
\begin{align}\label{Rnear}
R(x) = &x^{-i \frac{\varpi}{2}}  \left(\frac{x}{\tau}+1\right)^{i\frac{\varpi}{2}-ik} \nn \\
&{}_2F_1\left(\frac12 - \beta - ik,\, \frac12+\beta-ik;\, 1- i \varpi;\, -\frac{x}{\tau} \right)\, ,
\end{align}
where ${}_2F_1(a,\, b;\, c;\, z)$ is the hypergeometric function. In the intermediate region $\tau \times {\rm max}(1, \varpi) \ll x \ll 1$, we have Eq.~\eqref{Rnear} reducing to
\begin{align}\label{Rnear2}
R(x) = &\frac{\Gamma(1-i\varpi)\Gamma(-2\beta)\tau^{1/2+\beta - i\varpi/2}}{\Gamma(1/2-\beta-ik)\Gamma(1/2-\beta+ik-i\varpi)} x^{-\frac12-\beta} \nn\\
&+ (\beta \rightarrow -\beta).
\end{align}
In the far region, i.e. $x\ll \tau \times  {\rm max}(1, \varpi)$, Eq.~\eqref{eqRx} reduces to
\be
x^2 \frac{\d^2 R}{\d x^2} + 2x \frac{\d R}{\d x} + \left[ (\omega r_+ + k)^2 - \ell(\ell+1)\right]\, R = 0\,,
\ee
the solution of which can be written as
\begin{align}\label{Rfar}
R(x) &= N_1 \times (2\nu)^{\frac12-\beta} x^{-\frac12-\beta} e^{-\nu x} \nn \\
&{}_1 F_1\left(\frac12 - \beta - i k, \, 1-2\beta,\, 2 \nu x \right) + N_2 \times \left(\beta \rightarrow - \beta \right)\, ,
\end{align}
where $\nu \equiv i \omega r_+$. In the intermediate region, the far region solution Eq.~\eqref{Rfar} reduces to
\be\label{Rfar2}
R(x) = N_1 \times (2\nu)^{\frac12 - \beta}x^{-\frac12-\beta} + N_2 \times (\beta \rightarrow - \beta)\,.
\ee
Matching Eq.~\eqref{Rfar2} with Eq.~\eqref{Rnear2} in the overlap region, we find that
\begin{align}
&N_1(\beta) = \frac{\Gamma(1-i\varpi) \Gamma(-2\beta)}{\Gamma(1/2-\beta-ik)\Gamma(1/2-\beta +ik - i\varpi)}\tau^{\frac12+\beta - i \frac{\varpi}{2}} (2\nu)^{-\frac12+\beta} \, ,\nn\\
&N_2(\beta)=N_1(-\beta) \, .
\end{align}
Therefore, the solution at spatial infinity approaches
\begin{align}
R(x\rightarrow \infty) \rightarrow &\Bigg[ N_1(\beta) \times (2\nu)^{-ik} \frac{\Gamma(1-2\beta)}{\Gamma(1/2-\beta-ik)} \nn \\
&+ (\beta \rightarrow -\beta) \Bigg] x^{-1-ik} e^{\nu x} \nn\\
+&\Bigg[ N_1(\beta) \times (2\nu)^{ik} \frac{\Gamma(1-2\beta)}{\Gamma(1/2-\beta+ik)}  (-1)^{-\frac12+\beta+ik}  \nn \\
&+ (\beta \rightarrow -\beta) \Bigg] x^{-1+ik}e^{-\nu x} \, .
\end{align} 
The outgoing boundary condition requires $R(x\rightarrow \infty) \rightarrow e^{\nu x}$; thus,
\be
N_1(\beta) \times (2\nu)^{ik} \frac{\Gamma(1-2\beta)}{\Gamma(1/2-\beta+ik)}  (-1)^{-\frac12+\beta+ik} + (\beta \rightarrow -\beta) = 0\, ,
\ee
which determined the frequencies of the QNMs.

As we explained in the main text, a small perturbation in the potential effectively modified the boundary condition. In particular, we showed that, for a potential perturbation $\delta V = \epsilon \delta(x-x_V)$, the boundary condition to the leading order in $\epsilon$ is given by
\be\label{eqomega}
N_1(\beta) \times (2\nu)^{ik} \frac{\Gamma(1-2\beta)}{\Gamma(1/2-\beta+ik)}  (-1)^{-\frac12+\beta+ik} + (\beta \rightarrow -\beta)  = \epsilon \Delta (x_V) \, ,
\ee
where
\be
\Delta (x_V) = \frac{F_-(x_V)}{F_+(x_V) F_-'(x_V)- F_-(x_V) F_+'(x_V)}\, .
\ee
As $x\rightarrow \infty$, we have $F_\pm(x) \rightarrow  x^{-1\pm ik}e^{\mp \nu x}$ and, hence,
\be
\Delta(x\rightarrow \infty) \propto x^{1 - 2 i \omega r_+} e^{i \omega r_+ x} \, .
\ee
Equation~\eqref{eqomega} can be rewritten as
\begin{align}\label{eqomega2}
&\left[\frac{\Gamma\left(2\beta\right)}{\Gamma
\left(-2\beta\right)}\right]^2 \frac{\Gamma\left(\frac12-\beta-ik\right)\Gamma\left(\frac12-\beta+ik\right)\Gamma\left(\frac12 - \beta + ik - i \varpi \right)}{\Gamma\left(\frac12+\beta-ik\right) \Gamma\left(\frac12+\beta+ik\right) \Gamma\left(\frac12 + \beta + ik - i \varpi \right)}  \nn \\=&\left(2\nu \tau\right)^{2\beta} \left[1+  \frac{\epsilon \,\Delta(x_V)}{N_2(\beta) \times (2\nu)^{ik} \frac{\Gamma(1+2\beta)}{\Gamma(1/2+\beta+ik)}  (-1)^{-\frac12-\beta+ik} }\right].
\end{align}

In the absence of the perturbation, we can set $\epsilon=0$ in Eq.~\eqref{eqomega2}. Since $\tau \ll 1$ in the near-extremal case, we find the frequency is near the poles of $\Gamma\left(\frac12 + \beta + ik - i \varpi \right)$, namely
\be
\omega \simeq - i (\ell+n+1)\kappa_+ \, ,
\ee
where $n = 0,\, 1,\, 2,\, \cdots$. While in the presence of the potential perturbation, we find that $|\Delta(x)|$ is unbounded as $x$ approaches to infinity. 

\subsection{RNdS spacetime}

For RNdS spacetime, we further define
\be
c = \frac{r_+}{r_c-r_+}\, ,
\ee
and Eq.~\eqref{eqR} can be written as
\be
C_2 R''(x) + C_1 R'(x) + C_0 R(x)=0
\ee
where the prime denotes the derivative with respect to $x$, and
\begin{align}\label{eqRxRNdS}
C_2=& x(x+\tau)(1-c x) \left[1+ c (x+4-\tau)\right]\nn\\
C_1=&-4 c^2 x^3 - 12 c^2 x^2 +2\left[1+ c(4-\tau)-c^2(4-\tau)\tau\right] x \nn \\
& +\tau + c(4-\tau)\tau \nn \\
C_0=& \left[1+c(4-\tau) + c^2(6-4\tau+\tau^2)\right] \Bigg\{\ell(\ell+1) + \nn\\
&\frac{\omega^2 r_+^2(1+x)^4 \left[1+c(4-\tau) + c^2(6-4\tau+\tau^2)\right]}{x(x+\tau)(1-c x)\left[1+c(4-\tau)+c x\right]} \Bigg\}\, .
\end{align}
In the extremal limit, we have $\tau \rightarrow 0$. In the regime where $1-c x \ll 1$, Eq.~\eqref{eqRxRNdS} becomes
\be
(1- c x) R''(x) - c R'(x) + \left(c^2 j^2 + \frac{c^2 w^2}{1- c x}\right) R(x) = 0\, ,
\ee
where
\begin{align}
&j^2 = \frac{(1+4 c+6 c^2)}{2+4 c}\ell(\ell+1)\,,\quad w = \frac{\omega}{2 \kappa_c},\nn \\
&\kappa_c = \frac{c(1+2 c)}{(1+c)^2(1+4 c + 6 c^2) r_+}.
\end{align}
The general solution in this region can be written as
\begin{align}
R(x \rightarrow 1/c) =&A_{\rm o} \Gamma(1-2iw) {\rm J}_{-2 i w}\left(2j \sqrt{1- cx}\right) \nn \\
+& B_{\rm o}  \Gamma(1+2iw) {\rm J}_{2 i w}\left(2j \sqrt{1- cx}\right) \, ,
\end{align}
where ${\rm J}_n(z)$ is the Bessel function of the first kind. The outgoing boundary condition indicates $A_{\rm o} = 0$.  In the presence of the infinitesimal potential perturbation, the correction on $A_{\rm o}/B_{\rm o}$ is proportional to $\epsilon (1- c x_V)^{1+i w}$. In the near-extremal limit, we have $\omega \propto \tau/c$, while $\tau \rightarrow 0$. Hence, unless $c$ goes to zero faster than $\tau$, e.g., in the case of a RN black hole, we find that the correction on $A_{\rm o}/B_{\rm o}$ is bounded as $x_V \rightarrow 1/c$.

\end{document}